# Mapping causal patterns in crystalline solids

Chris Nelson,[1] Anna N. Morozovska,[2] Maxim A. Ziatdinov,[1,3] Eugene A. Eliseev,[4] Xiaohang Zhang,[5] Ichiro Takeuchi,[5] and Sergei V. Kalinin,[1,a]

[1] Center for Nanophase Materials Sciences, Oak Ridge National Laboratory, Oak Ridge, TN 37831

[2] Institute of Physics, National Academy of Sciences of Ukraine, 46, pr. Nauky, 03028 Kyiv, Ukraine

[3] Computational Sciences and Engineering Division, Oak Ridge National Laboratory, Oak Ridge, TN 37831

[4] Institute for Problems of Materials Science, National Academy of Sciences of Ukraine, Krjijanovskogo 3, 03142 Kyiv, Ukraine

[5] Department of Materials Science and Engineering, University of Maryland, College Park, MD 20742

The evolution of the atomic structures of the combinatorial library of Sm-substituted thin film $BiFeO_3$ along the phase transition boundary from the ferroelectric rhombohedral phase to the non-ferroelectric orthorhombic phase is explored using scanning transmission electron microscopy (STEM). Localized properties including polarization, lattice parameter, and chemical composition are parameterized from atomic-scale imaging and their causal relationships are reconstructed using a linear non-Gaussian acyclic model (LiNGAM). This approach is further extended toward exploring the spatial variability of the causal coupling using the sliding window transform method, which revealed that new causal relationships emerged both at the expected locations, such as domain walls and interfaces, but also at additional regions forming clusters in the vicinity of the walls or spatially distributed features. While the exact physical origins of these relationships are unclear, they likely represent nanophase separated regions in the morphotropic phase boundaries. Overall, we pose that an in-depth understanding of complex disordered materials away from thermodynamic equilibrium necessitates understanding not only of the generative processes that can lead to observed microscopic states, but also the causal links between multiple interacting subsystems.

[a] sergei2@ornl.gov

Foundational advances in condensed matter physics and materials science are inseparably linked with developments in characterization methods that provide progressively more detailed information on the structure and functionality of matter. The introduction of X-ray scattering methods by Bragg over a century ago enabled the first determinations of the atomic structure of solids and also stimulated the emergence of physical formalism based on reciprocal space descriptions, including zone theory, quasiparticles, and order parameter-based theories. In this case, observables such as structure factors or inelastic scattering can be directly matched to the theoretical descriptors. Comparatively, details on how these structures emerge were of lesser interest with the connections established via average defect concentrations and the potential presence and dynamics of extended defects.

Not surprisingly, this approach was limited for materials that lacked long-range translational symmetry including structural, dipole, and spin glasses.[1, 2] For these systems, the kinetic limitations during the formation process led to the formation of high defect densities and frustrated bond networks. Interestingly, systems lacking long-range order also emerge as a result of geometric frustration and symmetry mismatch between interacting subsystems and in principle correspond to (one of the) highly degenerate potential energy minima separated by high activation barriers. These materials have rapidly become one of the central topics of condensed matter physics due to both the fundamental physical interest and the unique functional properties they exhibit. [3-5] However, studying these materials via scattering methods has proven to be highly non-trivial and remains an active area of research.[6-8]

The emergence of atomically resolved scanning probe microscopy (SPM) over the last two decades of the twentieth century[9-11] and especially the rapid progress in aberration-corrected (scanning) transmission electron microscopy (STEM)[12-14] and atomic probe tomography (APT) during the last decade have made the observation of the atomic structures and compositions of 2D and small volume “bulk” materials routine. Consequently, materials structures and chemistries are becoming accessible for examination at the individual atom level (APT, tomographic reconstructions in STEM). In other cases, information on the atomic column positions and compositions averaged across the sample thickness is available. However, advances in visualizing the atomic structures necessitate equivalent progress in the development of a corresponding descriptive language that allows both parsimonious descriptions of observed structures (i.e., local equivalent of primitive cell) and can be used to build physical models.

In many cases, descriptions of atomic-level structures have been accomplished using macroscopic descriptors such as Fourier transforms, with the amplitude interpreted similar to scattering studies and phases used as additional information such as for geometric phase analysis in (S)TEM.[15-24] In several cases, this approach was adapted to the sliding window transform[25-28] to explore variability within an imaged region. Alternatively, identification of atomic-level features allows for strategies based on graph theory, molecular fragments, or other descriptors.[29, 30] The atomic and continuous descriptors can be further combined as demonstrated in Ref. [[31]]. Here, the use of the dimensionality reduction methods that allow for the incorporation of invariance enable parsimonious descriptions. However, these models invariably build a correlative model of the solid structure, in which the number of required descriptors can still be exceedingly large. For example, describing point defects in solids or potential atomic configurations in solid solutions far from a phase transition is relatively straightforward, whereas extended defects or materials in the vicinity of a phase transition require significantly larger and a potentially divergent number of descriptors.

The alternative approach for descriptions of solids can be based on generative models. For example, the description of a (potentially structurally complex) microstructure belonging to the Ising universality class requires only the knowledge of the corresponding exchange integrals. This approach directly underpins multiple areas of theoretical condensed matter and statistical physics, where the lattice models or force-fields form the basis for the description.[32] Interestingly, while the prediction of the structure and functionality from the theoretical models is by now extremely well developed, the inverse problem of reconstruction of the model parameters from experimental observations is only at the beginning, with only a few recent examples of generative model reconstructions.[33, 34]

However, the discovery of a generative model is not always sufficient to fully understand the materials physics. For example, in morphotropic ferroelectrics the dopants can pin polarization; however, polarization instabilities can in turn drive cation redistribution. Similarly, can we establish whether the average electron concentration or local chemistry drive distortion patterns in doped chalcogenides? Such questions can be answered by performing a macroscopic experiment. For example, the electron concentration effect and chemical environment can be separated by electrostatic gating in field effect devices. Similarly, ferroelectric instability can be tuned by pressure independent of composition. However, in many cases direct experiments are expensive or impossible, or are associated with additional confounding effects. For example, electrostatic gating to change the electron concentration often couples to chemical changes in a material. Hence, the question is whether the local observations of multiple spatially resolved degrees of freedom in STEM can be used to analyze the causal mechanisms that are operational in real materials.

Here, we explore causal interactions in cation-substituted multiferroic materials across the ferroelectric-non ferroelectric phase transition boundary from atomically resolved STEM measurements on the combinatorial library of Sm-doped $BiFeO_3$. A linear non-gaussian acyclic model (LiNGAM)[35] is used to reconstruct the causal chain and estimate the linear connections between the observed variables. This approach is further adapted to explore the spatial variability and concentration dependence of these interactions.

## 1. Causal descriptions in physics

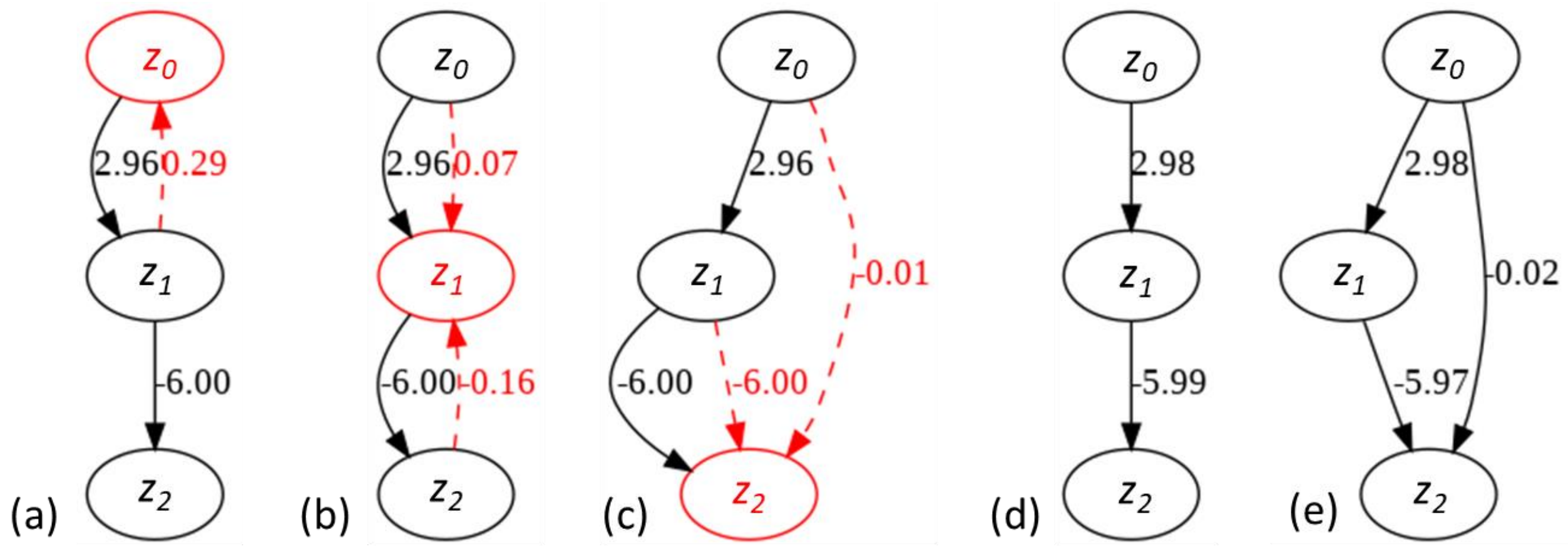

**Figure 1.** Regression vs. causal inference in a noisy toy model from system of linear equations 1a, and 1b. Shown are causal chain and regression coefficients for regression of (a) $z_0(z_1,z_2)$, (b) $z_1(z_0,z_2)$, and (c) $z_2(z_0,z_1)$. In comparison, LiNGAM reconstruction using (d) direct fit or (e) bootstrapping allows reconstruction of causal chain from observational data ($z_0$, $z_1$, $z_2$).

To illustrate the concepts involved in causal analysis, Figure 1 shows the simplified cause and effect chain between three variables, $z_0$, $z_1$, and $z_2$. As an example, $z_0$ can be the partial pressure of gas, $z_1$ is the concentration of oxygen vacancies, and $z_2$ is the conductivity of the material; however, similar examples can be chosen from any physical subfield. We assume that the relationship between these parameters is given by:

$$z_1 = 3\, z_0 + \varepsilon_1 \tag{1a}$$

$$z_2 = -6\, z_1 + \varepsilon_2 \tag{1b}$$

where $\varepsilon_1$ and $\varepsilon_2$ are random noise with a uniform distribution on (-1,1). We further assume that a set of observables ($z_0$, $z_1$, $z_2$) are available for observation. In the case when the causal physics of the process are known, the analysis of the data is straightforward. Regression of $z_1$ on $z_0$, $z_1 = z_1(z_0)$ yields the dependence of the vacancy concentration on gas pressure, whereas regression $z_2(z_1)$ yields the dependence of conductivity on vacancy concentration and $z_2(z_0)$ yields the dependence of conductivity of gas pressure. Note that implicit in this analysis is that there are no other active mechanisms; for example the temperature (that can affect both conductivity and vacancy concentration) is constant, there is no field effect gating of the materials, and there are no changes of doping level.

However, the situation is different when the physics are not known. Even if it is known that the relationships between the observables are linear, it is insufficient to establish the correct form of the functional dependence between them. In other words, any of the variables $z_0$, $z_1$, $z_2$ can be regressed on the remaining variables, as illustrated in Figure 1 (a-c). For regression $z_0 = z_0(z_1,z_2)$, $z_0$ depends on $z_1$ according to inverted Eq. (1a) and does not depend on $z_2$. However, while numerically correct, this relationship does not represent the physics of the process (i.e., the partial pressure of the gas does not depend on vacancy concentration). For regression $z_1 = z_1(z_0,z_2)$, the regression coefficients are close to zero and do not represent the physical relationship either. Only in case of the regression $z_2 = z_2(z_0,z_1)$ are the regression coefficients meaningful and close to the ground truth Eq. (1a,b), as shown in Fig. 1 (c). However, even in this case it is clear that based on the regression analysis only, we cannot answer questions such as "to what degree will the change of gas pressure affect the conductivity?" More fundamentally, from the data only it is not clear which of the cases (a-c) should be chosen.

This behavior is well-understood in statistics and illustrates the pitfalls of the statistical analysis of data without the physical model even for relatively trivial cases.[36, 37] This situation can be much more complicated for a larger number of variables, in cases where the relationships between the variables are more complex, i.e., the causal graph has a more complex structure, potentially including loops, relationships between variables that are non-linear, and there are external non-observed variables (confounders) that affect several of the observed variables. These considerations have stimulated the development of the statistical methods for analysis of the causal structure in the observational data, including the structure of the causal graph, directionality of the causal effects, presence of confounders, and linear and non-linear relationships between the variables. A recent overview of these methods is given by Mooji.[38]

Here, we adopt the linear non-Gaussian acyclic model (LiNGAM) proposed by Shimizu for analysis of time series data,[35, 39-43] building upon the independent component analysis (ICA) by Oja.[44, 45] LiNGAM specifically assumes that the relationships between the observed variables are linear, there are no external confounders, and the external influences are non-Gaussian, i.e., the observations can be represented as:

$$\boldsymbol{z} = \boldsymbol{A}\boldsymbol{z} + \boldsymbol{\varepsilon} \quad (2)$$

where $\mathbf{z}$ is the vector of observations, $\mathbf{A}$ is the adjacency matrix establishing the relationships between them, and $\boldsymbol{\varepsilon}$ is the noise vector. Under the assumption of the directed causal graph, the matrix, $\mathbf{A}$, has the lower diagonal shape. Correspondingly, LiNGAM seeks to find the linear decomposition of data via the ICA approach and then seeks to find the optimal representation of the link matrix via permutations.

For the toy model Eq. 1 (a,b), LiNGAM allows for the reconstruction of the causal graph between the variables. The reconstruction is shown in Fig. 1 (d,e) for the bootstrapped and direct solution, respectively. Note that the algorithm established the causal graph and linear relationship coefficients between the variables based on observational data only.

## 2. Causal relationships in $BiFeO_3$

We further explore the causal relationships in the $(Sm_xBi_{1-x})FeO_3$ (Sm-BFO) epitaxial thin film system grown via a combinatorial spread method via LiNGAM, extending previously reported pairwise causal analysis approach.[46] The growth details, preparation of the samples for STEM imaging, and initial data analysis and quantification are described in our previous publications (arXiv:2003.08575).[46, 47] The full data set is available as a part of the provided Jupyter notebook.

Here, we analyze the dependence of localized nominal physical properties as polarization, lattice parameter, and chemical composition from parameterization of atomic-scale HAADF-STEM images. For basic descriptors we choose parameters from the 5-cation local neighborhood that are sensitive to cation composition: the total intensity of A-cation columns (Bi,Sm), $I_{14}$, intensity of B-site columns (Fe), $I_5$, to the structure: lattice parameters, $a$ and $b$, unit-cell angle, α, and projected unit-cell volume, *Vol*, and to electrical polarization, components $P_x$ and $P_y$. derived from non-centrosymmetry between the A and B sublattices. This vector ($I_{14}$, $I_5$, $a$, $b$, α, *Vol*, $P_x$, $P_y$) is defined for each unit cell within field of view.

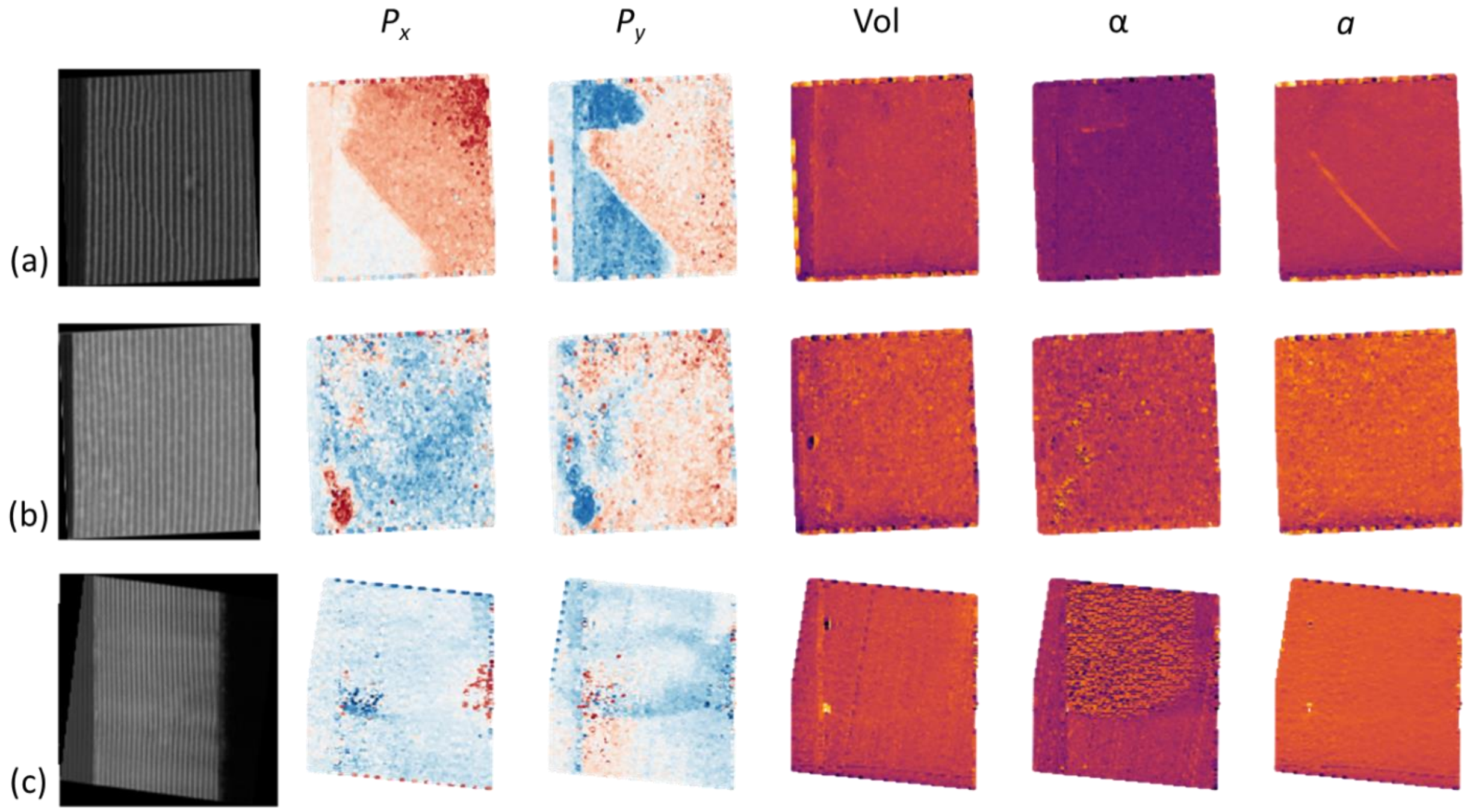

**Figure 2.** STEM images and partial set of order parameter maps for three Sm-BFO compositions explored (a) $x$ = 0, (b) $x$ = 0.07, and (c) $x$ = 0.2, corresponding to rhombohedral ferroelectric $BiFeO_3$, composition close to the morphotropic phase boundary, and orthorhombic antiferroelectric composition, respectively.

Shown in Figure 2 are STEM images and a partial set of the order parameter maps for three compositions of the Sm-BFO chosen in the Fig. 2 (a) ferroelectric region, Fig. 2 (b) at the morphotropic phase boundary, and in the Fig. 2 (c) orthorhombic phase. We note that the STEM images contain clearly visible spatial features including ferroelectric and substrate regions and domain walls. In addition, more subtle variations of contrast due to second phase inclusions, nanodomains, and surface preparation damage can be detected. Here, we aim to explore the causal relationship between the observables and further extend it for the analysis of spatially resolved features using the LiNGAM methods and codes developed by Shimizu (https://github.com/cdt15/lingam).

The LiNGAM method allows for incorporation of prior knowledge in the form of defined causal relationships. Here, we postulate that the column intensities $I_{14}$ and $I_5$ are independent variables, representing the combined effect of the sample thickness and assumption of the frozen cation composition. With this approximation, the causal graph depicting the relationship between the variables for three representative compositions is shown in Figure 3.

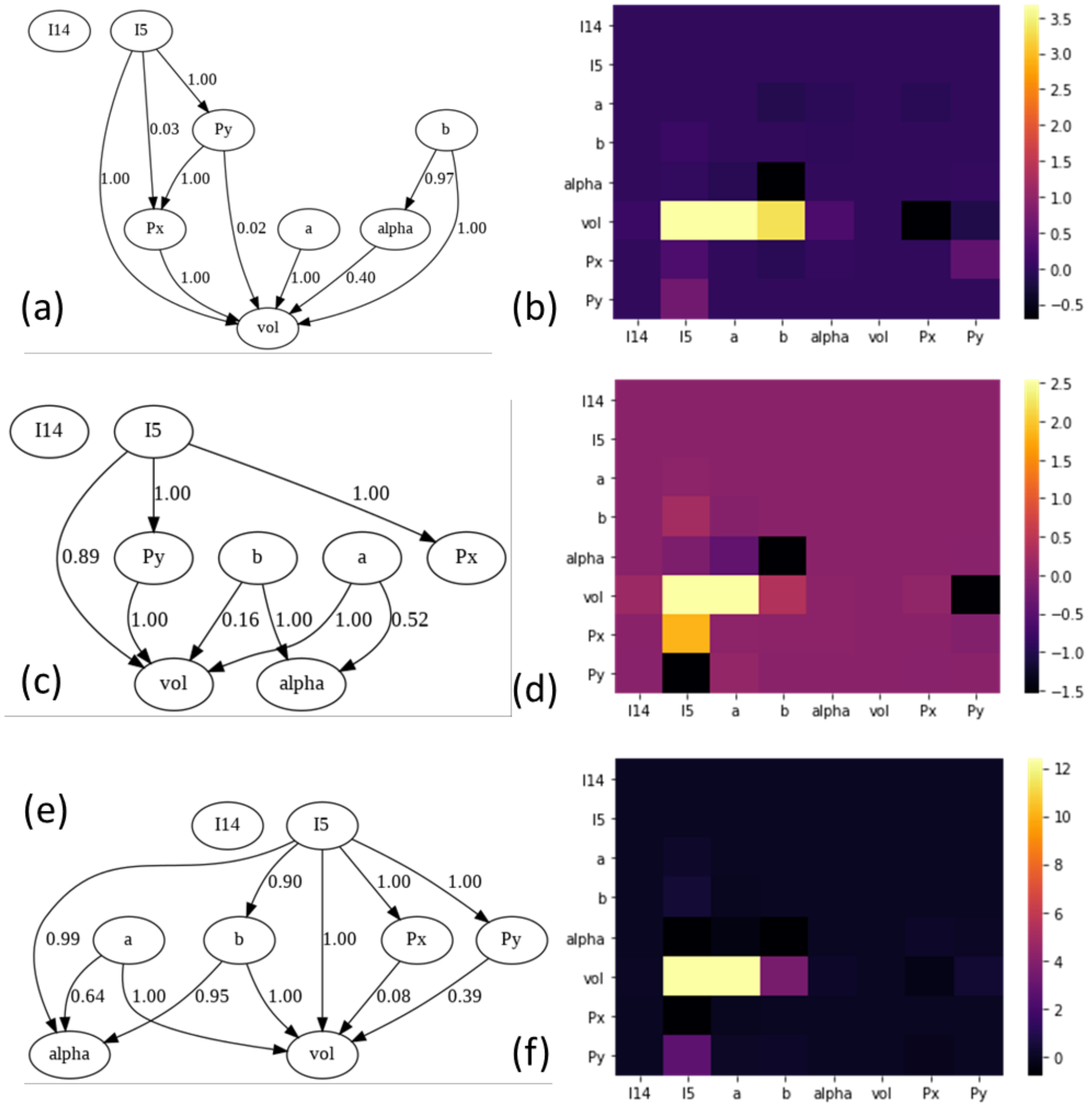

**Figure 3.** Causal graphs and adjacency matrices for (a,b) ferroelectric, $x = 0$, (c,d) morphotropic, $x = 0.07$, and (e,f) orthorhombic, $x = 0.2$ compositions.

From the data shown in Figure 3, it is clear that the $I_{14}$ variable does not causally affect any other variables. For the ferroelectric composition, $x = 0$, the control variables are $P_y$ and $b$ that control $P_x$ and $\alpha$, respectively, which in turn control the projected unit-cell volume, *Vol*. The graphs for the non-ferroelectric compositions are almost identical, with the lattice parameters and polarization components being independent variables and $\alpha$ and *Vol* being controlled. For the compositions at the morphotropic boundary and in the orthorhombic region, the causal graphs are almost identical with the polarization and unit-cell parameters affected by the chemical variable and in turn controlling the unit-cell *Vol* and $\alpha$. The corresponding linear regression coefficients in the form of adjacency matrices are shown in Figure 3 (b,d,f). For comparison, all maps are shown at the same scale. Note the strong similarity between the elements with the dominant coupling associated with the projected unit cell volume, *Vol*.

### 3. Sliding LiNGAM transform

We further extend this approach to explore the spatial variability in the thin films. To implement this approach, we extend the sliding window approach previously used in conjunction with Fourier or Radon transform/linear unmixing transformations.[25-27] Here, the image plane is subdivided into square tiles and the LiNGAM analysis is performed within each tile. For convenience, we chose the stride of the tiles to be half of the tile size so that adjacent regions overlapped by 50%. Similar to the analysis of the whole image, the input is the ($I_{14}$, $I_5$, $a$, $b$, α, $Vol$, $P_x$, $P_y$) vector within the tile and the output are the 8x8 adjacency matrices of the causal graph and the probability matrix.

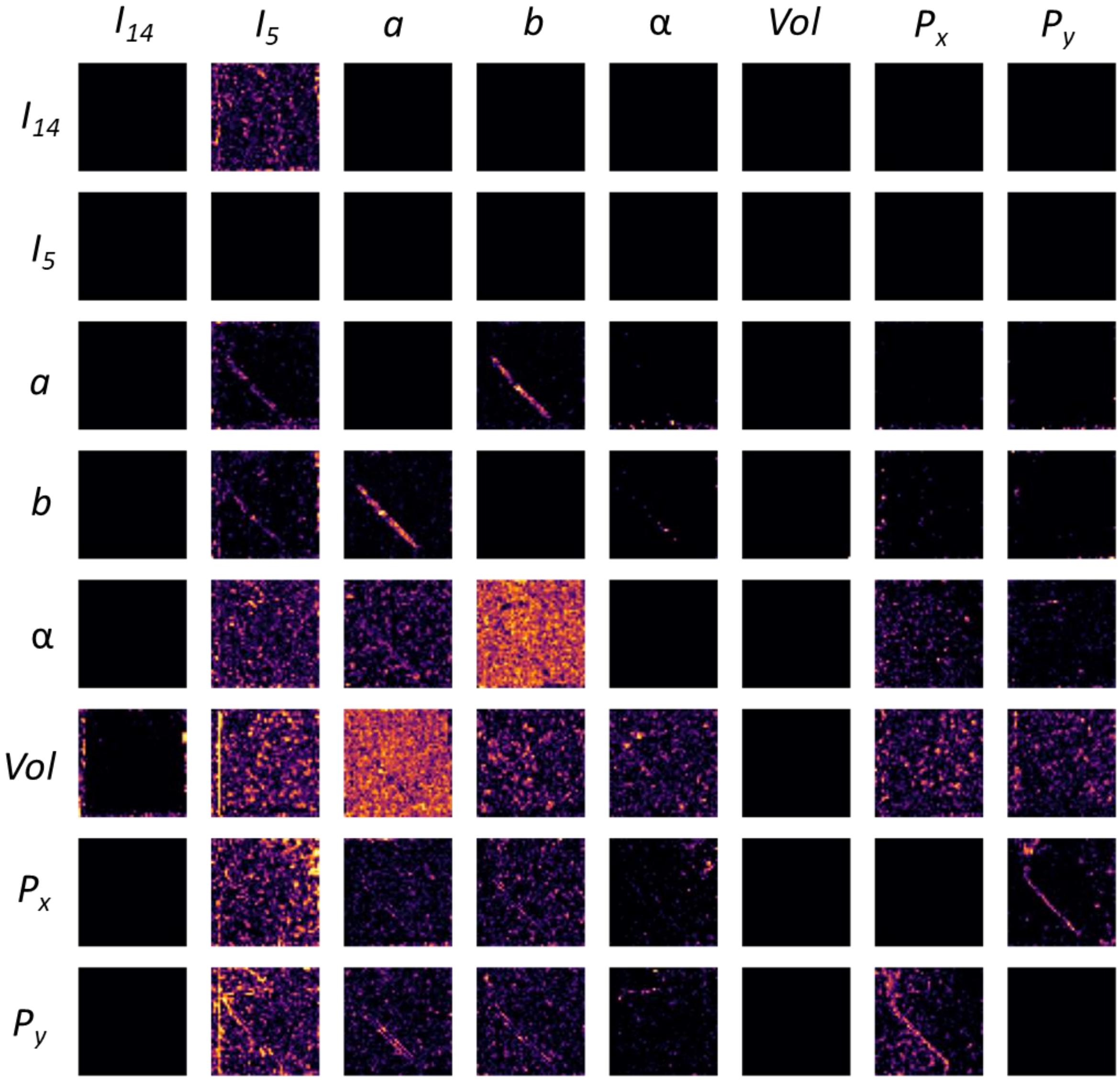

**Figure 4.** Sliding window LiNGAM transforms on the $Sm_xBi_{1-x}FeO_3$ data set for $x = 0$. Note that the use of bootstrapping method for several variables leads to both $x \rightarrow y$ and $x \leftarrow y$ relationships; for example, for lattice parameters ($a$, $b$) and polarization components ($P_x$, $P_y$). We pose that this behavior indicates that variables are linked via physical link, but this link is non-causal (or cyclic).

The sliding analysis of the Sm-BFO composition is shown in Figure 4. A simple examination of the images corresponding to the individual elements of the adjacency matrix illustrates that slightly over 50% are zero, corresponding to the absence of a ***local*** causal link between them. For several of the elements including $Vol \leftarrow a$, $Vol \leftarrow b$, and $\alpha \leftarrow b$ are almost uniform, without clearly visible structural elements. However, several of the maps exhibit visible structural features. For example, $Vol \leftarrow I_5$ and $P_y \leftarrow I_5$ maps clearly illustrate a feature at the interface between the film and the substrate. At the same time, $a \leftarrow b$ and $P_x \leftarrow P_y$ maps clearly illustrate feature associated with the domain wall. Note that these links are bi-directional. We interpret this behavior as an indicator of a non-causal link between these parameters

To gain further insight into the observed behaviors, we explore the spatial variability of the adjacency matrix using dimensionality reduction. Previously, these methods were extensively used for the analysis of hyperspectral imaging data in electron and scanning probe microscopies[48-53] and were later been extended to sliding transform approaches. In principle, this can be implemented using linear methods such as principal component analysis (PCA), non-negative matrix factorization (NMF), or non-linear methods including Gaussian processing,[54, 55] fully connected and variational autoencoders, or graph networks. Here, we choose PCA as the simplest method to perform optimal decomposition in the information theory sense, i.e., with the components ranked in order of variability.

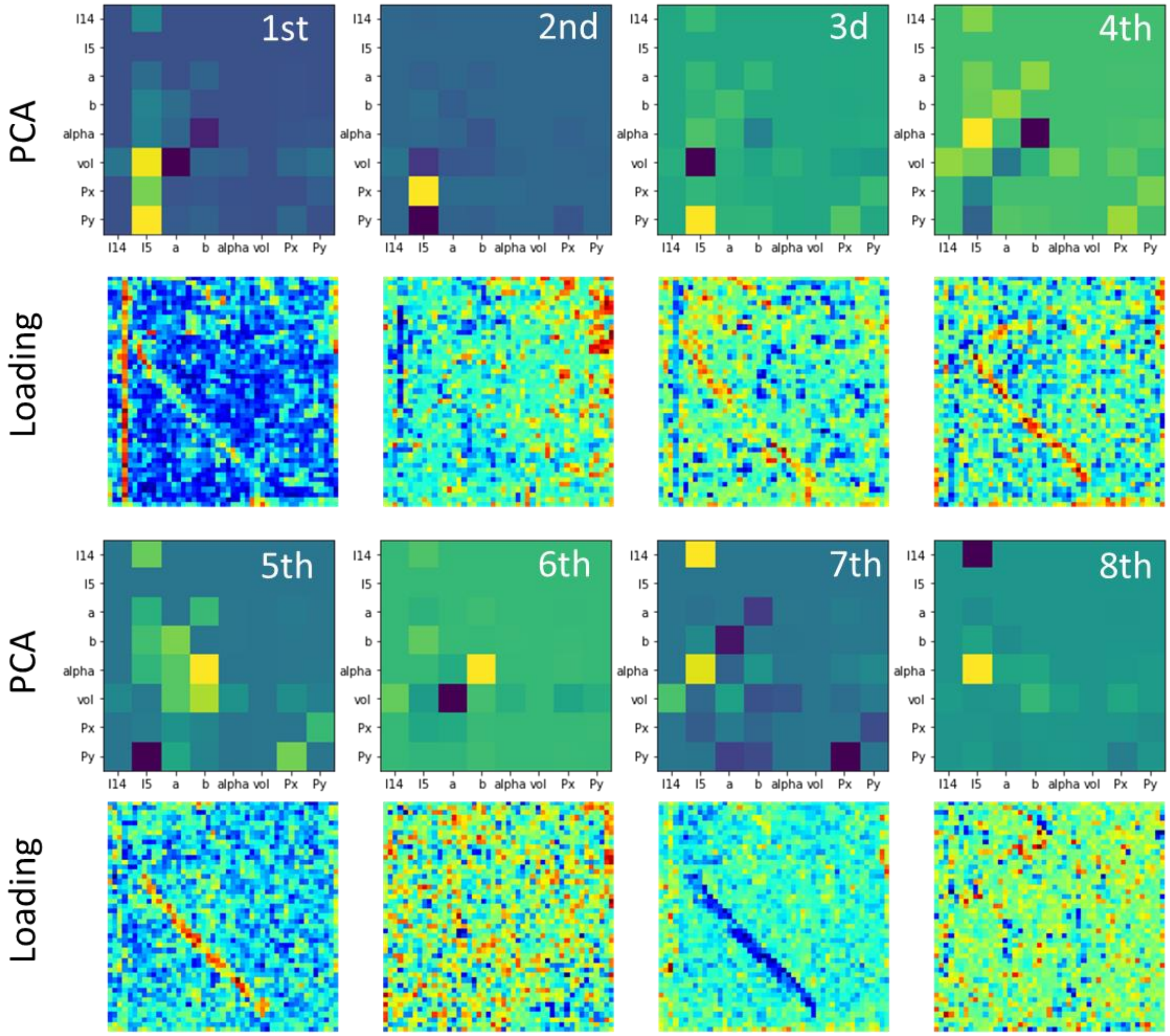

**Figure 5.** PCA analysis of LiNGAM adjacency matrices for $BiFeO_3$. Shown are first 8 components and corresponding loading maps. First components contain a clearly visible feature localized at film-substrate interface, whereas 4$^{th}$, 5$^{th}$, and 7$^{th}$ components clearly delineate domain wall.

The PCA analysis for pure $BiFeO_3$ is shown in Figure 5. Here, it is immediately obvious that the loading map corresponding to the first component exhibits variability at the interface between the film and the substrate, which is generally absent or much weaker in other loading maps. The corresponding component matrix is dominated by coupling between $I_5$ and the polarization and projected volume as well as between the volume and in-plane lattice parameter, *a*. The 4$^{th}$, 5$^{th}$, and 7$^{th}$ components exhibit visible features at the ferroelectric domain wall. The corresponding components are dominated by the couplings $I_5$- α, α -b, and $I_{14}$-$I_5$, respectively.

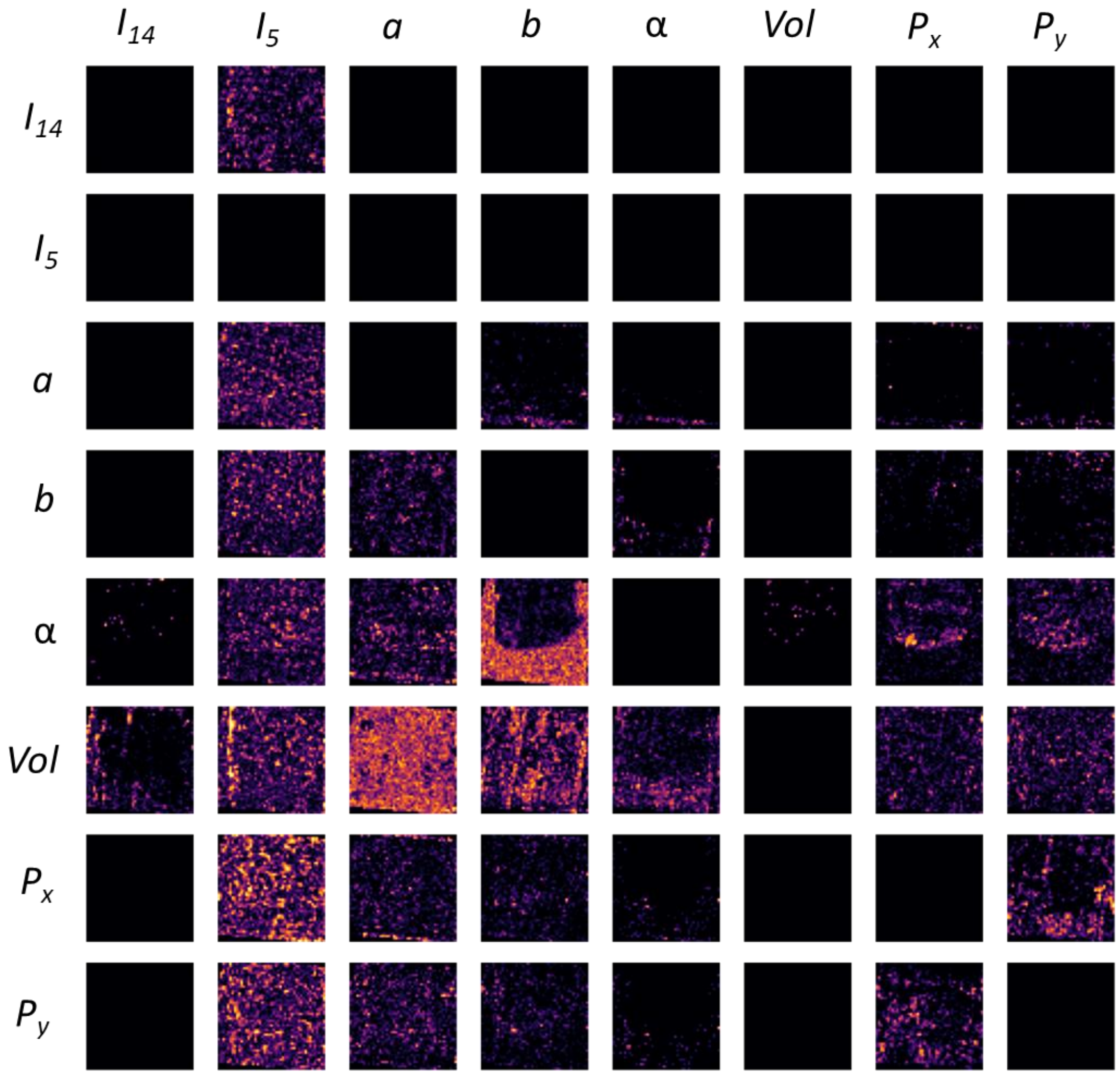

**Figure 6.** Sliding window LiNGAM transform on Sm-BFO data set for $x$ = 0.2 (orthorhombic phase).

The sliding LiNGAM transform for the 20%-doped Sm-BFO ($x$ = 0.2) is shown in Figure 6. Similar to other compositions, here, the majority of the maps are either zero, indicating the absence of a causal link between the descriptors, or are spatially uniform. However, several of the elements exhibit non-trivial dynamics. Similar to other compositions, the $Vol \leftarrow I_5$ element visualizes a clearly visible feature at the domain wall. The $\alpha \leftarrow b$ map clearly visualizes the orthorhombic domains. Interestingly, the $\alpha \leftarrow P_x$ and $\alpha \leftarrow P_y$ maps illustrate spatial features associated with the wall between orthorhombic domains. Examination of the sliding transform for other explored compositions further reveals the presence of spatially ordered structures in some of the maps, suggesting the potential of this method to reveal internal structural relationships within the material.

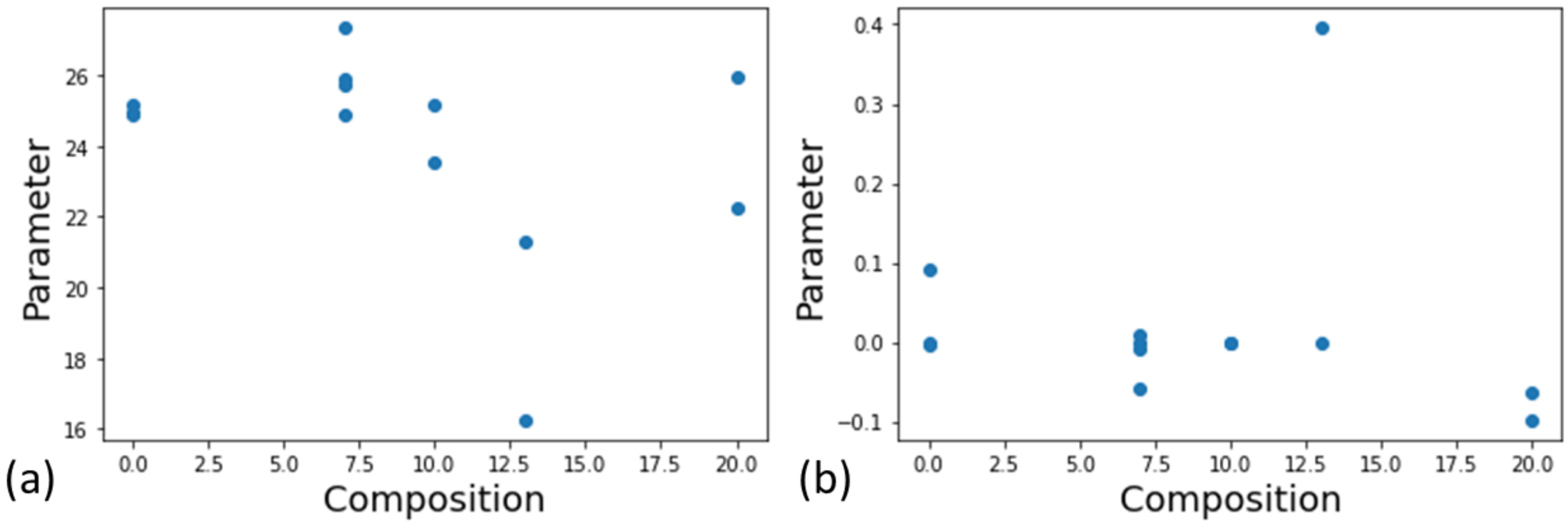

**Figure 7**. Concentration dependence of (a) $Vol \leftarrow a$ and (b) $P_x \leftarrow P_y$ couplings.

Finally, we attempted to analyze the concentration dependence of the adjacency matrix, i.e., the coupling strength between the observables. In this case, the analysis was performed over 13 independent STEM images corresponding to four different concentrations acquired over multiple days. In general, given the strong spatial variability within the data set, we do not expect the statistics to be sufficient to discern the finer details of the concentration dependence; however, we expect the overall trends to be detectable. Examination of all 64 elements of the adjacency matrix demonstrated that most of these elements are either zero or are constant. This suggests the universality of the corresponding couplings across the composition series. Non-trivial behaviors were observed for the $Vol \leftarrow a$ and $P_x \leftarrow P_y$ couplings, as shown in Figure 7. For $Vol \leftarrow a$ we observe the minimum at the morphotropic phase boundary (MPB) composition. At the same time, $P_x \leftarrow P_y$ shows the effective change of sign from positive to negative at the MPB.

## 4. Linear model of BFO

To explain these observations, we explore the relationships between the observables descriptors via the thermodynamic Landau-Ginzburg -Devonshire (LGD) theory. In LGD, the state of the system can be found via minimization of the free energy functional for defined global constraints. Here, we derive the local representation of LGD equations, linearize them around the ground states, and represent the relationship in the form of Eq. (1) with the observables in the STEM experiment being the state vector.

In the most general case, description of the rare-earth Sm-doped $BiFeO_3$ requires antiferrodistortive (AFD), ferroelectric (FE), and antiferroelectric (AFE) long-range order.[56-60]

This necessitates two vectorial long-range order parameters, polarization components, and oxygen octahedral tilts. The bulk part of LGD thermodynamic potential consists of the following contributions:

$$G = \int d^3x\left(\Delta G_{AFD} + \Delta G_{BQC} + \Delta G_{FE} + \Delta G_{EST} + \Delta G_{EL}\right). \qquad (3a)$$

The compact form of the AFD contribution is given in Ref.[61]. Operating at room temperature, we will assume that the main impact of the oxygen octahedral tilt is the renormalization of other coefficients in Eq. (1) due to the AFD-FE biquadratic coupling, $\Delta G_{BQC}$. Since the oxygen octahedral tilts are hardly visible by the current realization of STEM, this circumstance allows us to factor out the non-observable variables from further consideration.

The compact form of the FE and AFE contributions are:

$$\Delta G_{FE} = a_i(P_i^2 + A_i^2) + a_{ij}\left(P_i^2P_j^2 + A_i^2A_j^2\right) + a_{ijk}\left(P_i^2P_j^2P_k^2 + A_i^2A_j^2A_k^2\right) + \gamma_{ij}^{ab}\left(P_iP_j - A_iA_j\right) + g_{ijkl}^{aa}\left(\frac{\partial P_i}{\partial x_k}\frac{\partial P_j}{\partial x_l} + \frac{\partial A_i}{\partial x_k}\frac{\partial A_j}{\partial x_l}\right) + g_{ijkl}^{ab}\left(\frac{\partial P_i}{\partial x_k}\frac{\partial P_j}{\partial x_l} - \frac{\partial A_i}{\partial x_k}\frac{\partial A_j}{\partial x_l}\right), \qquad (3b)$$

where the FE and AFE order parameters, $P_i = \frac{1}{2}\left(P_i^a + P_i^b\right)$ and $A_i = \frac{1}{2}\left(P_i^a - P_i^b\right)$, respectively, are introduced, $P_i^a$ and $P_i^b$ are the polarization components of two equivalent sublattices, "$a$" and "$b$", respectively.[62] Numerical values of the phenomenological coefficients $a_i$, $a_{ij}$, $a_{ijk}$, and gradient coefficients, $g_{ij}$, included in Eq. (2b) are mostly known for pure $BiFeO_3$ and can be found in **Suppl. Mat.**, **Table SI**.

The electrostriction contribution is:

$$\Delta G_{EST} = -Q_{ijkl}^{aa}\sigma_{ij}(P_kP_l + A_kA_l) - Q_{ijkl}^{ab}\sigma_{ij}(P_kP_l - A_kA_l), \qquad (3c)$$

where $\sigma_{ij}$ are the components of elastic stress tensor. Electrostriction coefficients are $Q_{ijkl}^{aa}$ and $Q_{ijkl}^{ab}$. The elastic and flexoelectric contributions are:

$$\Delta G_{EL} = -\frac{1}{2}s_{ijkl}\sigma_{ij}\sigma_{kl} - \frac{1}{2}F_{ijkl}\left(\sigma_{ij}\frac{\partial P_k}{\partial x_l} - P_k\frac{\partial \sigma_{ij}}{\partial x_l}\right) + V_{ij}\sigma_{ij}N_d. \qquad (3d)$$

Here, $s_{ijkl}$ are the components of the elastic compliance tensor and $F_{ijkl}$ are the flexoelectric tensor components. The last term in Eq. (3d) is the chemical expansion due to the appearance of elastic defects, i.e., R-impurity with concentration, $N_d$, characterized by the Vegard strain tensor, $V_{ij}$,[63, 64] the value of which depends on the impurity and, as a rule, varies in the range (–5 – +5) $10^{-29}$ $m^3$. The full form of expressions (3) depends on the concrete form of the parent phase symmetry.

For a spatially uniform (i.e., domain-free) system, the equilibrium equations for the long-range order parameters can be found as partial derivatives of Eq.(3a), $\frac{\partial G}{\partial P_i} = 0$ and $\frac{\partial G}{\partial A_i} = 0$. These equations should be solved along with the equation of state for the elastic stress, $\frac{\partial G}{\partial \sigma_{ij}} = -u_{ij}$, where $u_{ij}$ is an elastic strain tensor.

For the general case of an inhomogeneous (e.g., domain structured or/and spatially modulated) system, one should solve the coupled Euler-Lagrange equations of state, which are expressed via the variational derivatives of the functional (3a):

$$\frac{\delta G}{\delta P_i} = -E_i, \qquad \frac{\delta G}{\delta A_i} = 0. \qquad (4a)$$

The electric field, $E_i$, can be found from electrostatic equations with boundary conditions at the surfaces, interfaces and/or electrodes. Elastic fields, which are, in fact, the secondary order parameters, satisfy the equation of state and mechanical equilibrium equations:

$$\frac{\partial G}{\partial \sigma_{ij}} = -u_{ij}, \qquad \frac{\partial \sigma_{ij}}{\partial x_j} = 0. \qquad (4b)$$

The strains (or stresses) should be defined at the system boundaries. Coupled equations (4), which give us the relations between the primary and secondary order parameters, are listed in their explicit form in **Suppl. Mat.**

To link the standard thermodynamic descriptors to the observables, we use the relationships between the in-plane lattice constants, $a$, $b$, and angle, $\alpha$, between them:

$$a = a_0(1 + u_{11}), \quad b = b_0(1 + u_{22}), \quad cotan(\alpha) = u_{12}. \tag{5}$$

Here, $a_0$ and $b_0$ are the bulk values of undoped BFO at a given temperature. The projected volume of 2D-unit cell is $V = ab|sin(\alpha)|$. The linearized changes of the unit-cell projected volume can be expressed via the 2D-trace of the strain tensor:

$$V = ab|sin(\alpha)| \approx a_0 b_0 (1 + u_{11} + u_{22}). \tag{6}$$

For high-resolution (HR)-STEM the observable variables in Eqs. (5)-(6) are the in-plane components of the FE order parameters, $P_1$ and $P_2$, lattice constants, $a$, $b$, $cos(\alpha)$, and changes of the projected unit cell volume, which in principle allow us to extract information about the joint action of the Vegard strain and electrostriction coupling. However, other components of the order parameters cannot be observed by HR-STEM imaging, which is why we need to find the simple (often approximate) linear relations between the observable and non-observable variables.

To do this, we linearized Eqs. (4) in the vicinity of spontaneous values, denoted by subscript "S", which are:

$$P_i = P_{Si}(\vec{x}) + \delta P_i, \qquad A_i = A_{Si}(\vec{x}) + \delta A_i, \tag{7a}$$

$$u_{ij} = u_{ij}^{S}(\vec{x}) + \delta u_{ij}, \quad \sigma_{ij} = \sigma_{ij}^{S}(\vec{x}) + \delta\sigma_{ij}. \tag{7b}$$

The magnitudes of the spontaneous values, $P_S$ or $A_S$, depend on the average concentration of Sm dopant atoms and the dependence can be modeled for the simplest linear model[65] as:

$$P_S = P_0\sqrt{1-\epsilon}, \qquad A_S = P_0\sqrt{1+\epsilon}, \tag{8}$$

where $P_0$ is the spontaneous polarization of the bulk pure BFO (at a given temperature) and we introduced the dimensionless doping factor $\epsilon$:

$$\epsilon(y) \cong \epsilon_0(y - y_0). \tag{9}$$

Here, $y$ is the average Sm concentration in the studied sample, $y_0$ is a characteristic concentration corresponding to the MPB, and the FE and AFE phases have the same Curie temperatures. The coordinate dependence of the spontaneous values, $P_{Si}(\vec{x})$ and $A_{Si}(\vec{x})$, is related with mesoscale domain walls, while $\delta P_i(\vec{x})$ and $\delta A_i(\vec{x})$ are nanoscale fluctuations. Here, $\vec{x} = \{x_1, x_2\}$, is the in-plane coordinate vector.

Linearized Eqs. (3)-(4) can be written in the form of a 3x3 block-matrix containing nine equations:

$$\begin{pmatrix} a_{ij}^{P} - g_{ijkl}^{P}\frac{\partial^2}{\partial x_k \partial x_l} & 0 & 0 \\ 0 & a_{ij}^{A} - g_{ijkl}^{A}\frac{\partial^2}{\partial x_k \partial x_l} & 0 \\ q_{inj}^{P} + F_{injl}\frac{\partial}{\partial x_l} & q_{inj}^{A} & -1 \end{pmatrix} \begin{pmatrix} \delta P_j \\ \delta A_j \\ \delta u_{in} \end{pmatrix} = \begin{pmatrix} \delta E_i + Q_{mjki}^{P} P_{Sk} \delta\sigma_{mj} \\ Q_{mjki}^{A} A_{Sk} \delta\sigma_{mj} \\ V_{in}\delta N_d - s_{injm}\delta\sigma_{jm} \end{pmatrix}, \tag{10a}$$

where the block-matrix elements are

$$a_{ij}^{P} = 2(a_i + 3a_{il}P_{Sl}^2 + 5a_{ilm}P_{Sl}^2 P_{Sm}^2)\delta_{ij} + \left(\gamma_{ij}^{ab} - Q_{mlji}^{P}\sigma_{ml}^{S}\right), \tag{10b}$$

$$a_{ij}^{A} = 2\left(a_i + 3a_{ij}A_{Sj}^2 + 5a_{ijk}A_{Sj}^2 A_{Sk}^2\right)\delta_{ij} - \left(\gamma_{ij}^{ab} + Q_{mlji}^{A}\sigma_{ml}^{S}\right), \tag{10c}$$

$$q_{inj}^{P} = 2Q_{inkj}^{P}P_{Sk}, \quad q_{inj}^{A} = 2Q_{inkj}^{A}A_{Sk}. \tag{10d}$$

The system (10) can be treated further analytically if we neglect the gradient terms, while the assumption is not well-grounded everywhere. However, it can be valid if the coordinate dependence can be regarded as smooth and then ascribed to the domain wall profiles of $P_{Si}(\vec{x})$ and $A_{Si}(\vec{x})$, while the fluctuations, $\delta P_i\ (\vec{x})$ and $\delta A_i\ (\vec{x})$, are characterized by a short-range gradient with almost a zero average. In this case, the antipolar order parameter can be expressed via other inobservable stress variations as $\delta A_n = \tilde{a}^A_{ni} Q^A_{mjki} A_{Sk} \delta\sigma_{mj}$, where $\tilde{a}^A_{ni} a^A_{ij} = \delta_{nj}$. Next, we exclude the non-observable variables from the left-hand side of Eqs. (10a) by re-designation of the right-hand side:

$$e^*_i = \delta E_i + Q^P_{mjki} P_{Sk} \delta\sigma_{mj} - a^P_{i3}\,\delta P_3, \quad i = 1{,}2, \tag{11a}$$

$$s^*_{ij} = V_{in}\delta N_d + \left(s_{injm} + q^A_{imj}\tilde{a}^A_{ni'} Q^A_{mjki'} A_{Sk}\right)\delta\sigma_{jm} + q^P_{ij3}\delta P_3, \quad i,j = 1{,}2. \tag{11b}$$

The truncated matrix form is:

$$\begin{pmatrix} a^P_{11} & a^P_{12} & 0 & 0 & 0 \\ a^P_{21} & a^P_{22} & 0 & 0 & 0 \\ -q^P_{111} & -q^P_{112} & 1 & 0 & 0 \\ -q^P_{121} & -q^P_{122} & 0 & 1 & 0 \\ -q^P_{221} & -q^P_{222} & 0 & 0 & 1 \end{pmatrix} \begin{pmatrix} \delta P_1 \\ \delta P_2 \\ \delta u_{11} \\ \delta u_{12} \\ \delta u_{22} \end{pmatrix} = \begin{pmatrix} e^*_1 \\ e^*_2 \\ s^*_{11} \\ s^*_{12} \\ s^*_{22} \end{pmatrix}, \tag{12}$$

Note that the in-plane strains in Eq. (12) are observable variables, namely $\delta u_{11} = \frac{a}{a_0} - 1$, $\delta u_{12} = cos(\alpha) \approx \frac{\pi}{2} - \alpha$, and $\delta u_{22} = \frac{b}{b_0} - 1$. Here, the projected volume, $V \approx ab$, since $cos(\alpha) \approx \frac{\pi}{2} - \alpha$ is extremely small. Assuming zero spontaneous stresses and na m3m parent symmetry, the elements of the block-matrix (12) are:

$$a^P_{11} = a^P_{22} \approx 2a_1[1 - \epsilon(y)] + 6a_{1l}P^2_{Sl}(\vec{x}) + \gamma^{ab}_{11}, \qquad a^P_{12} = a^P_{21} = \gamma^{ab}_{12} = \gamma^{ab}_{21}, \tag{13a}$$

$$q^P_{111} = 2Q^P_{11}P_{S1}(\vec{x}), \quad q^P_{112} = 2Q^P_{12}P_{S2}(\vec{x}), \quad q^P_{121} = 2Q^P_{44}P_{S2}(\vec{x}), \tag{13b}$$

$$q^P_{122} = 2Q^P_{44}P_{S1}(\vec{x}), \quad q^P_{221} = 2Q^P_{12}P_{S1}(\vec{x}), \quad q^P_{222} = 2Q^P_{11}P_{S2}(\vec{x}). \tag{13c}$$

From this analysis, only the polarization dependence is expected to yield terms dependent on concentration, in agreement with observational studies.

We further proceed to explore possible mechanisms of spatial variability of observed behaviors. The coordinate dependence of the spontaneous values, $P_{Si}\ (\vec{x})$, related with a mesoscale domain wall in the direction, *x*, perpendicular to the wall plane, $x = 0$, can be fitted by the following expression:

$$P_S(x) \cong \sqrt{-\frac{a_1}{a_{11}}}\sqrt{1 - \epsilon(y)}\,tanh\left(\frac{x}{L^P_C}\right). \tag{14}$$

Being repetitive, here, $\epsilon(y) \cong \epsilon_0(y - y_0)$, $y$ is the average Sm concentration in the studied sample, $y_0$ is a characteristic concentration corresponding to the MPB, and the FE and AFE phases have the same Curie temperature. For a Sm-doped BFO, the value $y_0$ varies in the range (15 – 25)% and the constant $\epsilon_0$ must be fitted to experimental data. Parameters used in the LGD calculations for AFD-FE perovskite BFO are listed in **Table SI**; however, the AFE-FE coupling constants, $\gamma^{ab}_{ij}$, are unknown. The correlation lengths, $L^P_C$, in a rhombohedral ferroelectric are listed in **Table SII** in **Suppl. Mat.**

For the sake of simplicity, we assume that $\delta\sigma_{mj} = 0$ and $\delta P_3 = 0$ when plotting Fig. 8. This immediately leads to $e^*_i = \delta E_i$, meaning that these non-observable variables are the components of local electric fields, and to $s^*_{ij} = V_{in}\delta N_d$, meaning that these non-observable variables are Vegard strains. If we also assume that $\gamma^{ab}_{11} = 0$, then $a^P_{11} = a^P_{22} = 2a_1[1 - \epsilon(y)]\left(1 - 3tanh^2\left(\frac{x}{L^P_C}\right)\right)$ and thus, $a^P_{11} = a^P_{22} = -4a_1[1 - \epsilon(y)] > 0$ far from the domain wall, while $a^P_{11} = a^P_{22} = 2a_1[1 - \epsilon(y)] < 0$ at the wall.

Figure 8 illustrates how the dimensionless ratio, $\delta P_2/\delta P_1$, behaves under a change of the random field ratio, $e_2^*/e_1^*$, which is independent of the other observable variables, $\delta u_{ij}$ (if the above assumptions are valid). Actually, from Eqs. (12) we obtained the simple expressions:

$$\delta P_1 = \frac{a_{11}^P e_1^* - a_{12}^P e_2^*}{\left(a_{11}^P\right)^2 - \left(a_{12}^P\right)^2}, \qquad \delta P_2 = \frac{a_{11}^P e_2^* - a_{12}^P e_1^*}{\left(a_{11}^P\right)^2 - \left(a_{12}^P\right)^2}, \qquad \frac{\delta P_2}{\delta P_1} = \frac{a_{11}^P (e_2^*/e_1^*) - a_{12}^P}{a_{11}^P - a_{12}^P (e_2^*/e_1^*)}. \tag{15}$$

From Eqs. (15a) and Figs. 8 we observe that the dependence of $\delta P_2/\delta P_1$ on $e_2^*/e_1^*$ reaches a maximum at the domain wall plane and the height of the maximum gradually decreases with an increase in $\epsilon(y)$. That said, the maximum at the wall is highest at the MPB, $\epsilon(y) = 0$. The sharpness of the maximum is virtually independent of $\epsilon(y)$. The ratio, $\delta P_2/\delta P_1$, becomes much more sensitive to the ratio, $e_2^*/e_1^*$, as $\epsilon(y)$ increases. The limit, $\epsilon(y) \rightarrow 1$, corresponds to the AFE phase.

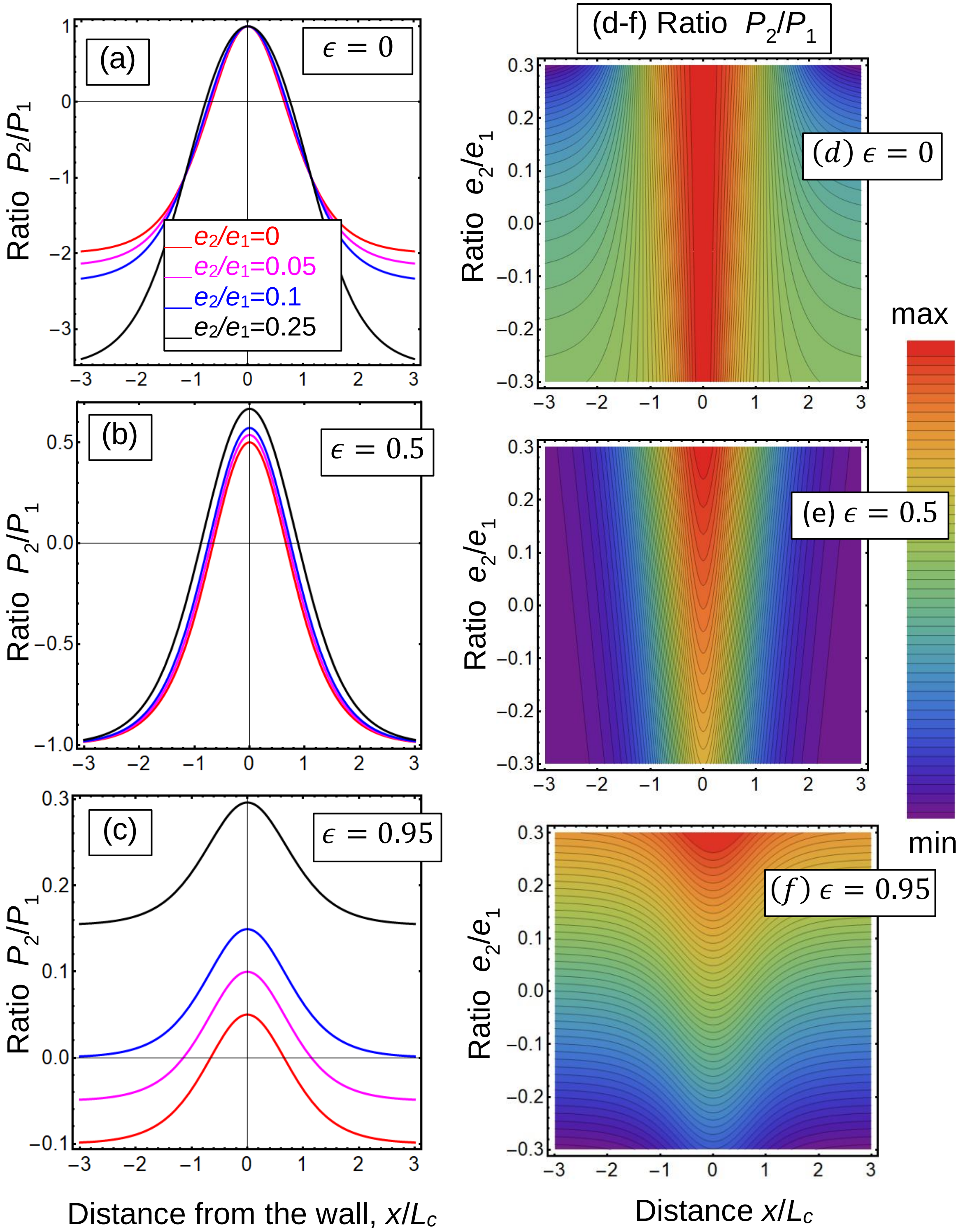

**Figure 8**. Ratio of observable variables, $\delta P_2/\delta P_1$, dependence on distance, $x/L_C^P$, from 180° domain wall calculated for different ratios of non-observable local fields, $e_2^*/e_1^*$, and different parameters, $\epsilon(y) = 0, 0.5, 0.95$ proportional to Sm concentration, $y$. **(a-c)** – linear plots, **(d-f)** – color maps. Red corresponds to maximal positive values, violet corresponds to minimal negative values. Other parameters, $a_{12}^P = a_{21}^P = -2a_1$.

Note that the strains, $\delta u_{ij}$, are linearly dependent on the observable variables, $\delta P_i$, and seemingly independent of one another (while can be dependent via $\delta P_i$). Actually, from Eqs.(12):

$$\delta u_{11} = q^P_{111}\delta P_1 + q^P_{112}\delta P_2 + s^*_{11},$$
$$\delta u_{12} = q^P_{121}\delta P_1 + q^P_{122}\delta P_2 + s^*_{12},$$
$$\delta u_{22} = q^P_{221}\delta P_1 + q^P_{222}\delta P_2 + s^*_{22}. \quad (16a)$$

For the case $P_{S1}(\vec{x}) = P_S(\vec{x})$ and $P_{S2}(\vec{x}) = 0$, corresponding to a single 180° domain wall in the x-direction, we obtain from Eqs.(13) that $q^P_{111} = 2Q^P_{11}P_S(\vec{x})$, $q^P_{112} = 0$, $q^P_{121} = 0$, and $q^P_{122} = 2Q^P_{44}P_S(\vec{x})$, $q^P_{221} = 2Q^P_{12}P_S(\vec{x})$, $q^P_{222} = 0$, and so in the case of in-plane lattice parameters, the angle between them and the projected volume are the following:

$$\frac{a}{a_0} = 1 + 2Q^P_{11}P_S(\vec{x})\delta P_1 + s^*_{11}, \quad \frac{b}{b_0} = 1 + 2Q^P_{12}P_S(\vec{x})\delta P_1 + s^*_{22}. \quad (16b)$$

$$\frac{\pi}{2} - \alpha = 2Q^P_{44}P_S(\vec{x})\delta P_2 + s^*_{12}, \quad \frac{V}{a_0 b_0} \approx 1 + 2(Q^P_{11} + Q^P_{12})P_S(\vec{x})\delta P_1 + s^*_{11} + s^*_{22}. \quad (16c)$$

From the last expressions, $\frac{a}{a_0} - 1 \sim \delta P_1$ and $\frac{b}{b_0} - 1 \sim \delta P_1$, but $\frac{\pi}{2} - \alpha \sim \delta P_2$.

Figure 9 illustrates the ratio of the observable variables, $\delta u_{11}/\delta P_1$, $\delta u_{12}/\delta P_2$, $\delta u_{22}/\delta P_1$, and $\delta V/\delta P_1$ in their dependence on the distance, $x/L^P_C$, from the 180° domain wall calculated for the different parameters, $\epsilon(y) = 0, 0.5, 0.95$, proportional to the Sm concentration, $y$. We put $a^P_{12} = a^P_{21} = -2a_1$ and $s^*_{ij} = 0$.

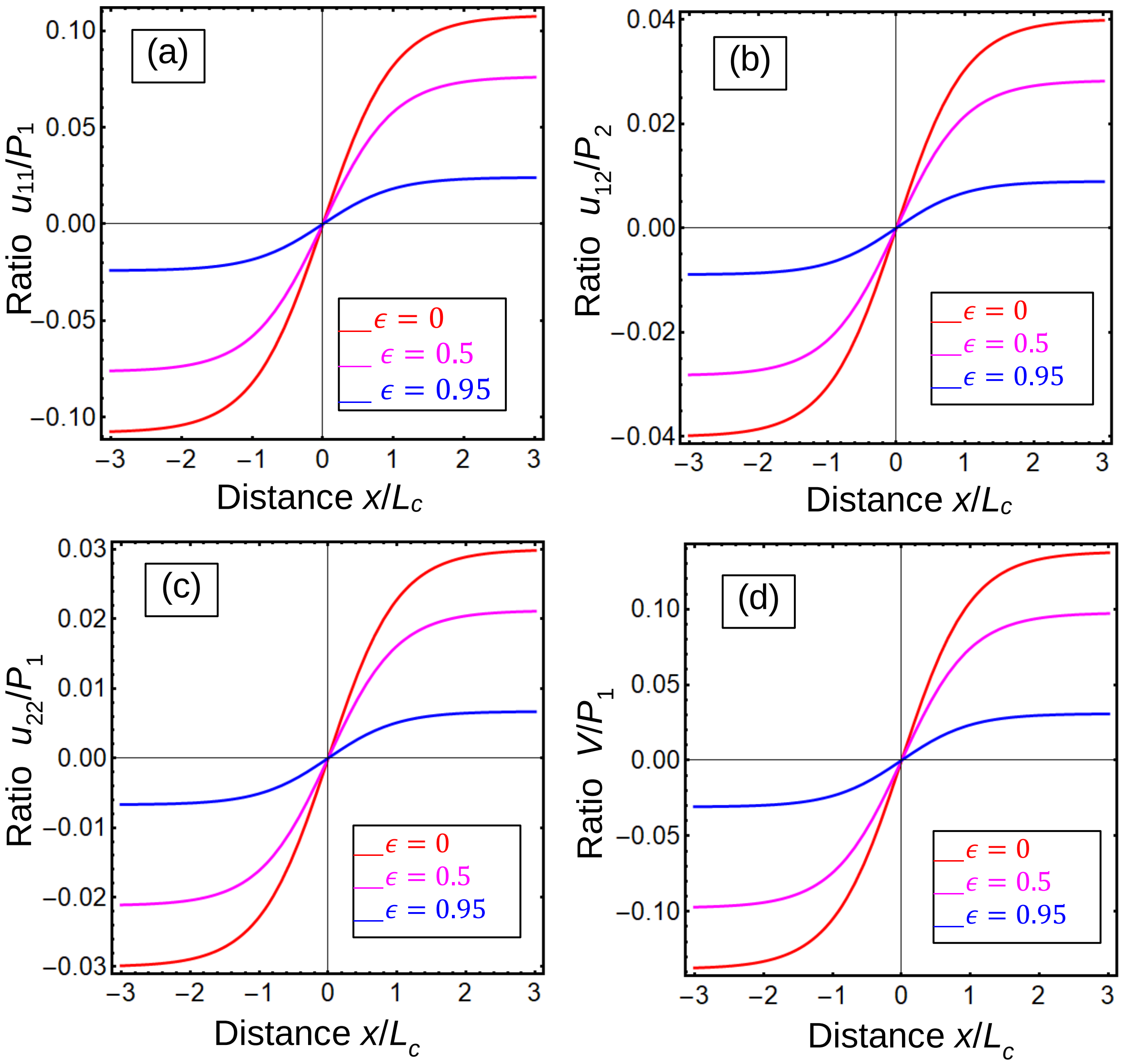

**Figure 9**. Ratio of observable variables (a) $\delta u_{11}/\delta P_1$, (b) $\delta u_{12}/\delta P_2$, (c) $\delta u_{22}/\delta P_1$, and (d) $\delta V/\delta P_1$ in their dependence on distance, $x/L_C^P$, from 180° domain wall calculated for different parameters, $\epsilon(y) = 0, 0.5, 0.95$, proportional to Sm concentration, $y$. Other parameters $a_{12}^P = a_{21}^P = -2a_1$, $s_{ij}^* = 0$.

## 5. Summary

To summarize, here we explore the causal linear coupling between polarization, lattice parameter, and chemical composition from experimentally observed single unit-cell level behaviors in Sm-doped $BiFeO_3$. This analysis suggests that the causal links between the observables is nearly universal for all Sm-concentrations/compositions. The variation of coupling strength between the different materials is generally similar across the phase transition boundary, with concentration-dependent behaviors observed for polarization only. This is directly confirmed by the linearized model based on the LGD theory for ferroelectrics.

We further explored the spatial variability of the causal coupling using the sliding window transform method. Interestingly, this approach revealed new causal relationships emerging at both expected locations, such as domain walls and interfaces, but also at regions in the vicinity of the walls or spatially distributed features where clusters formed. While the exact physical origins of these relationships are unclear, they likely represent nanophase separated regions in morphotropic phase boundaries.

Overall, we pose that an in-depth understanding of complex disordered materials necessitates understanding of not only generative processes that can lead to observed microscopic states, but also the causal links between the observed variables. While generative models are sufficient to describe systems in the state of thermodynamic equilibrium, non-equilibrium systems may require causal descriptors that can describe cause and effect relationships between multiple interacting subsystems.

**Acknowledgements:** This effort (ML and STEM) is based upon work supported by the U.S. Department of Energy (DOE), Office of Science, Basic Energy Sciences (BES), Materials Sciences and Engineering Division (S.V.K., C.T.N.) and was performed and partially supported (M.Z.) at the Oak Ridge National Laboratory's Center for Nanophase Materials Sciences (CNMS), a U.S. Department of Energy, Office of Science User Facility. A.N.M. work is supported by the National Academy of Sciences of Ukraine (the Target Program of Basic Research of the National Academy of Sciences of Ukraine "Prospective basic research and innovative development of nanomaterials and nanotechnologies for 2020 - 2024", Project № 1/20-H, state registration number: 0120U102306) and received funding from the European Union's Horizon 2020 research and innovation programme under the Marie Skłodowska-Curie grant agreement No 778070. The work at the University of Maryland was supported in part by the National Institute of Standards and Technology Cooperative Agreement 70NANB17H301 and the Center for Spintronic Materials in Advanced infoRmation Technologies (SMART) one of centers in nCORE, a Semiconductor Research Corporation (SRC) program sponsored by NSF and NIST. The authors gratefully acknowledge Dr. Karren More (ORNL) for careful reading and editing of the manuscript.